\documentclass[letter]{aa}
\usepackage{amssymb,amsmath}
\usepackage{graphicx}
\usepackage{xcolor,natbib}
\usepackage{multirow}
\usepackage[T1]{fontenc}
\usepackage{ae,aecompl}
\usepackage{newtxtext, newtxmath}
\usepackage{float}
\usepackage{subfigure}
\usepackage{longtable}
\usepackage{enumitem}
\usepackage{longtable,listings}
\usepackage[flushleft]{threeparttable}
\usepackage{parcolumns}
\usepackage{hyperref}

\def\n2{[N~{\sc ii}]$\lambda6583$\AA}
\def\si{[S~{\sc ii}]$\lambda6716$\AA}
\def\sii{[S~{\sc ii}]$\lambda6732$\AA}
\def\o3{[O~{\sc iii}]$\lambda5007$\AA}
\def\obj{SDSS J1011+5442}

\begin{document}

\title{Clues for the accretion regulated dust torus in the changing-look AGN SDSS 
J101152.98+544206.4}

\titlerunning{Evolving dust torus in CLAGN}

\author{XueGuang Zhang}

\institute{Guangxi Key Laboratory for Relativistic Astrophysics, School of Physical Science 
and Technology, GuangXi University, No. 100, Daxue Road, Nanning, 530004, 
P. R. China \ \ \ \email{xgzhang@gxu.edu.cn}}

\abstract
{Dust torus plays the key role in determining active galactic nuclei (AGN) observational 
appearance. Here, the scenario of accretion regulated central dust torus is tested for the 
first time in the individual changing-look AGN (CLAGN) SDSS J1011+5442. Through the 
dependence of broad H$\alpha$ luminosity on continuum luminosity, the scenario of moving dust 
clouds can be ruled out in SDSS J1011+5442. Meanwhile, virial BH mass in the bright state is 
consistent with the M-sigma relation determined mass, indicating the virialization assumptions 
efficient in central BLRs. However, the virial BH mass determined in the dim state is 60 times 
smaller than the M-sigma relation determined value. The contrary properties of broad H$\alpha$ 
in different states can be naturally explained by the scenario of accretion regulated dust 
torus. Below a critical Eddington ratio, opening angle of dust torus declines with increasing 
accretion rate, leading to only outer part of central BLRs for broad H$\alpha$ with smaller 
line widths detected in the dim state but all the BLRs detected in the bright state. The 
results in this manuscript not only indicate properties of central dust torus having apparent 
effects on variability properties of CLAGN, but also indicate that studying CLAGN could 
provide further clues to check dynamical evolving models for dust torus in AGN.}

\keywords{
galaxies:active - galaxies:nuclei - quasars: supermassive black holes}

\maketitle

\section{Introduction}

	SDSS J101152.98+544206.4 (=SDSS J1011+5442) at redshift 0.246 has been reported as a 
changing-look AGN (CLAGN) as discussed in \citet{run16, lw25}, due to its optical AGN type 
being changed from 1 to 1.9 and then to 1 from 2003 to 2015 and then to 2024. Furthermore, 
after analyzing both photometric and spectroscopic variability properties, the scenario of 
variations in intrinsic accretion rates has been accepted in SDSS J1011+5442 as discussed 
in \citet{lw25}, rather than the scenarios of effects of moving dust clouds \citep{lc15, 
rb16} and tidal disruption events \citep{ta19, zh21, wl24} applied in the CLAGN \citep{rt23}.

	However, there is one point which cannot be explained or expected by the scenario of 
variations in intrinsic accretion rates in the CLAGN SDSS J1011+5442, after checking 
properties of broad Balmer emission lines. If accepted virialization assumptions in central 
broad emission line regions (BLRs) \citep{ve02, pe04, gh05, sh11, mt22}, in SDSS J1011+5442 
from bright state to dim state, the broad H$\alpha$ has its line width being decreased, 
contrary to expectations by the virialization assumptions.

	Furthermore, accepted the scenario of variations in intrinsic accretion rates in SDSS 
J1011+5442, variations of accretion rates should regulate spatial structures of central dust 
torus, as the receding torus model in \citet{at05, ar11, mg16, mi19, bc21} and the 
radiation-regulated model in \citet{zh18, ra22, as25}. The changes in spatial structures of 
central dust torus could probably lead different parts of central BLRs to be detected in 
different states, to explain the unique variability properties of broad emission lines in 
CLAGN. The individual CLAGN SDSS J1011+5442 with apparent variations in central accretion 
rates provides the best chance to check the changes in spatial structures of central dust 
torus, which is the main objective of the manuscript.

	The manuscript is organized as follows. Section 2 presents the spectroscopic results 
and main discussions of the CLAGN SDSS J1011+5442. Main conclusions are given in Section 3. 
Throughout the manuscript, we have adopted the cosmological parameters of 
$H_{0}$=70 km s$^{-1}$ Mpc$^{-1}$, $\Omega_{m}$=0.3, and $\Omega_{\Lambda}$=0.7.

\begin{figure*}
\centering\includegraphics[width = 18cm,height=3.8cm]{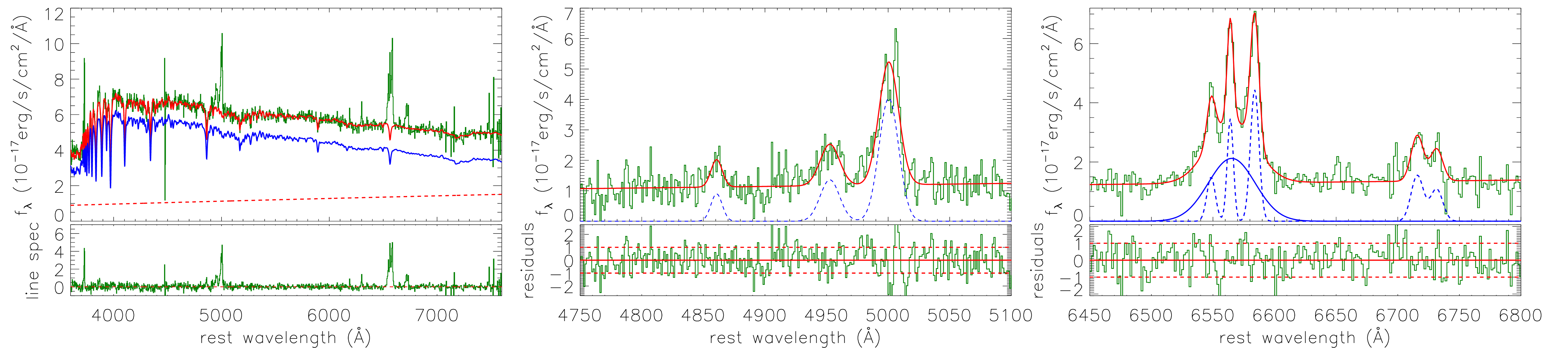}
\caption{Spectroscopic properties of \obj~ in dim state. Top left panel shows the SDSS 
spectrum (solid line in dark green) and the determined best descriptions to the host galaxy 
contributions (solid blue line) and AGN continuum emissions (dashed red line). Solid red line 
shows the sum of host galaxy contributions and AGN continuum emissions. Bottom left panel 
shows the spectrum after subtractions of host galaxy contributions and AGN continuum emissions, 
with horizontal dashed red line marking intensities to be zero. Top middle panel shows the 
spectrum around H$\beta$ (solid line in dark green) and the best fitting results (solid red 
line), after subtractions of host galaxy contributions. Dashed blue lines show the determined 
Gaussian components to the narrow H$\beta$ and [O~{\sc iii}] doublet. Bottom middle panel 
shows the residuals calculated by the shown spectrum minus the best fitting results and then 
divided by uncertainties of the shown spectrum, with horizontal solid and dashed lines in 
red marking residuals=$0,\pm1$. Right panels show the results to emission lines around 
H$\alpha$, similar as those in the middle panels. In top right panel, dashed blue lines show 
the determined Gaussian components in narrow emission lines, and solid blue line shows the 
determined broad H$\alpha$.}
\label{dime}
\end{figure*}

\begin{figure*}
\centering\includegraphics[width = 18cm,height=3.8cm]{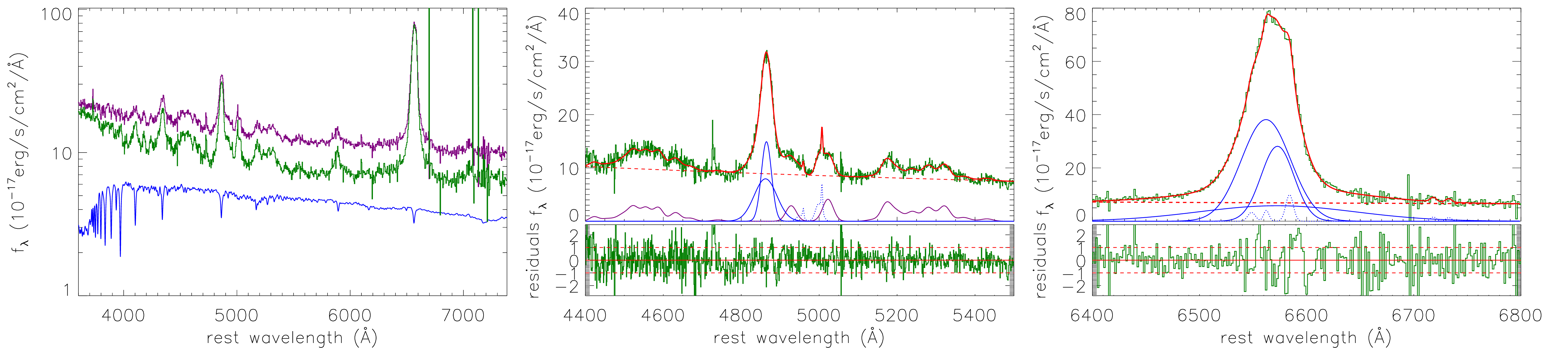}
\caption{Spectroscopic properties of \obj~ in bright state. Left panel shows the SDSS spectrum 
(solid purple line) in bright state. Solid blue line shows the host galaxy contributions 
determined through the spectrum in dim state. Solid line in dark green shows the component 
calculated by the SDSS spectrum in the bright state minus the host galaxy contributions. Top 
middle panel shows the best fitting results (solid red line) to the emission lines around 
H$\beta$, after subtractions of the host galaxy contributions. Dashed blue lines show the 
determined Gaussian components to the narrow H$\beta$ and [O~{\sc iii}] doublet, solid purple 
lines show the determined optical Fe~{\sc ii} emissions, solid blue lines show the determined 
Gaussian components in broad H$\beta$. Dashed red line shows the determined AGN continuum 
emissions underneath the emission lines. Bottom middle panel shows the corresponding residuals, 
with horizontal solid and dashed lines in red marking residuals=$0,\pm1$. Right panels show 
the results to the emission lines around H$\alpha$ in the bright state, after subtractions 
of the host galaxy contributions. Dashed blue lines show the determined Gaussian components 
in narrow H$\alpha$ and [N~{\sc ii}], [S~{\sc ii}] doublets, solid blue lines show the 
determined Gaussian components in broad H$\alpha$. Dashed line in red shows the determined 
AGN continuum emissions underneath the emission lines.}
\label{bright}
\end{figure*}

\section{Spectroscopic results and discussions}

	The SDSS spectrum of \obj~ (plate-mjd-fiberid=8181-57073-0827) in dim state is shown 
in top left panel of Fig.~\ref{dime}, with apparent stellar absorption features. Before 
measuring emission line properties, host galaxies contributions are firstly determined and 
subtracted. Here, the commonly accepted simple stellar population (SSP) method discussed 
in \citet{bc03, ka03, cm05, cm17} has been applied. Similar as what we have recently done 
in \citet{zh24a, zh24b, gz25, zh25}, sum of the strengthened, shifted and broadened 39 SSPs 
discussed in \citet{bc03, ka03} are applied to describe the host galaxy contributions. 
Meanwhile, a power law function is applied to describe the intrinsic AGN continuum emissions, 
due to weak but apparent broad H$\alpha$ emissions. Then, with emission lines being masked 
out, through the Levenberg-Marquardt least-squares minimization technique (the MPFIT package) 
\citet{mc09}, the host galaxy contributions and AGN continuum emissions can be determined 
and shown in top left panel of Fig.~\ref{dime}. Meanwhile, the stellar velocity dispersion 
(the broadening velocity of the SSPs) can be measured as 211$\pm$17km/s, consistent with 
the SDSS provided value 205$\pm$30km/s, indicating the measured stellar velocity dispersion 
to be reliable enough.

\begin{figure}
\centering\includegraphics[width = 9cm,height=4.75cm]{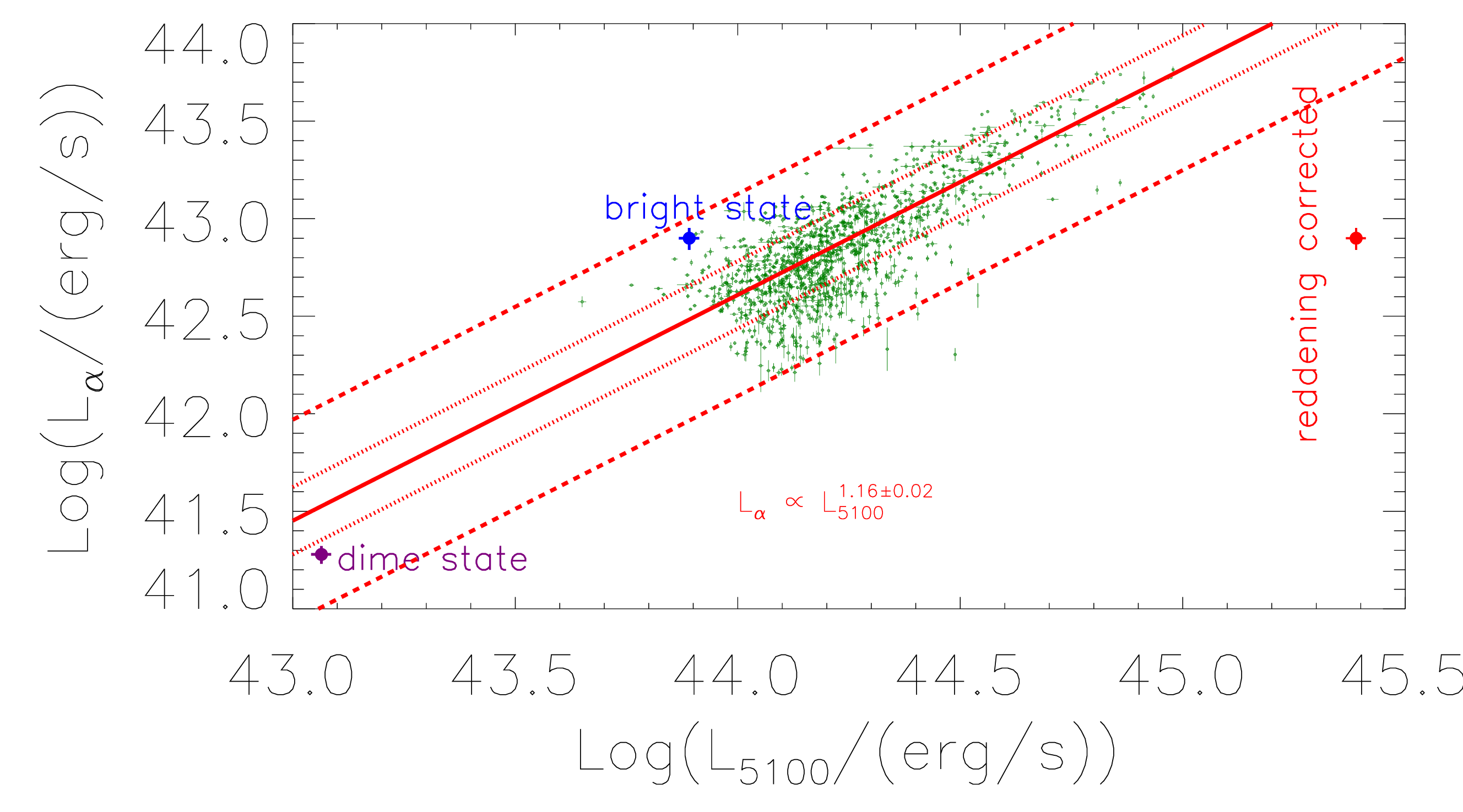}
\centering\includegraphics[width = 9cm,height=4.75cm]{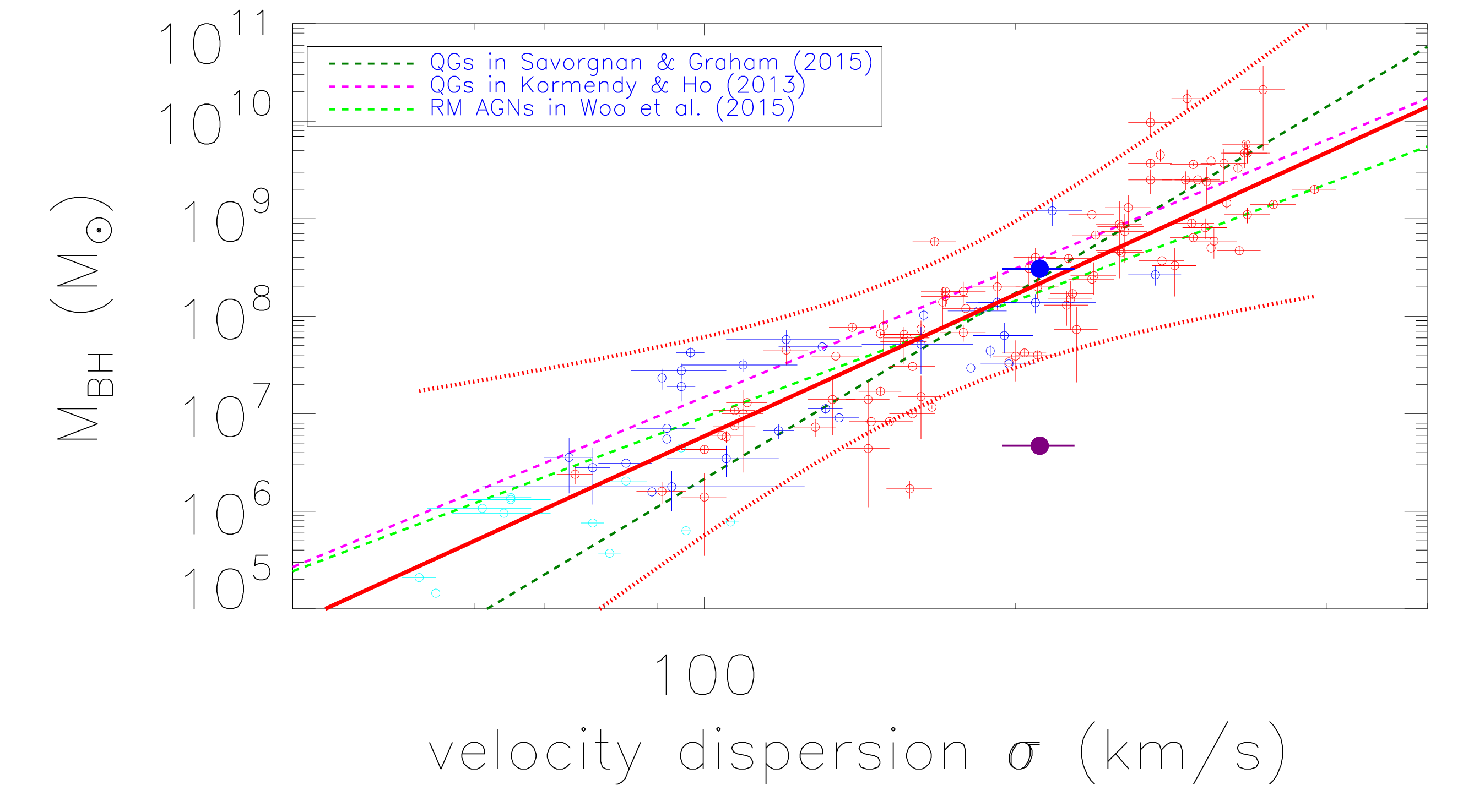}
\caption{Properties of \obj~ in space of $L_{\alpha}$ versus $L_{5100}$ (top panel) and of 
BH mass versus stellar velocity dispersion (bottom panel). In top panel, small blue dots 
with error bars show the results for the selected 1158 unobscured SDSS quasars from 
\citet{sh11}, solid, dotted and dashed lines in red show the best fitting results 
($L_\alpha\propto L_{5100}^{1.16\pm0.02}$) to the dependence and the corresponding 1RMS and 
3RMS scatters. Solid circle plus error bars in purple and in blue show the results of \obj~ 
in dim state and in bright state, respectively. In bottom panel, open circles plus error 
bars in red, in blue and in cyan show the results for the quiescent galaxies in \citet{sg15} 
and the reverberation mapped broad line AGN in \citet{wy15} and the tidal disruption events 
in \citet{zl21}. Solid line and dashed lines in red show the best fitting results to the 
dependence and the corresponding 5$\sigma$ confidence bands. Dashed lines in dark green, 
in purple, in green show the determined results to describe the dependence from \citet{sg15, 
kh13, wy15}. Solid circle plus error bars in purple and in blue show the results of \obj~ in 
dim state and in bright state, respectively.}
\label{cl}
\end{figure}

	After subtractions of the host galaxy contributions in the dim state, emission lines 
around H$\beta$ (rest wavelength from 4750 to 5100\AA) and around H$\alpha$ (rest wavelength 
from 6450 to 6800\AA) in the line spectrum can be measured by multiple Gaussian components. 
For each narrow emission line, one Gaussian function with intensity not smaller than zero is 
applied. For each broad Balmer emission line, one Gaussian function is applied with emission 
intensity not smaller than zero and line width not smaller than 400km/s. Meanwhile, for 
[O~{\sc iii}]$\lambda4959,5007$\AA~ doublet ([N~{\sc ii}]$\lambda6549,6583$\AA~ doublet), 
the same redshift, the same line width in velocity space and the flux ratio 3:1 have been 
accepted for the applied Gaussian components. Here, [O~{\sc i}]$\lambda6300,6363$\AA~ doublet 
are not considered, due to their very weak emission intensities. And a power law function 
is applied to describe the AGN continuum emissions underneath the emission lines around 
H$\beta$ (H$\alpha$). Then, through the Levenberg-Marquardt least-squares minimization 
technique, the best fitting results ($\chi^2/dof\sim0.91$) and corresponding residuals to 
the emission lines in the dim state are shown in the middle and right panels of 
Fig.~\ref{dime}. The residuals are calculated by the line spectrum minus the best fitting 
results and then divided by the uncertainties of the line spectrum. The measured parameters 
and corresponding $1\sigma$ uncertainties of the Gaussian emission components are listed in 
Table~1 in the Appendix A. The best fitting results can be applied to confirm that the \obj~ 
is a Type-1.9 AGN in the dim state, due to no broad H$\beta$ but apparent broad H$\alpha$ 
after accepted the common AGN-type classifications as listed in Table 1 in \citet{rc16}.

	Meanwhile, the SDSS spectrum of \obj~ (plate-mjd-fiberid=0945-52652-0022) in the 
bright state is shown in the left panel of Fig.~\ref{bright}. Accepted the host galaxy 
contributions determined in the dim state, the line spectrum in the bright state can be 
simply determined by the bright spectrum minus the host galaxy contributions. Then, similar 
multiple Gaussian components are applied to describe the emission lines with rest wavelength 
from 4400 to 5600\AA~ and from 6400 to 6800\AA. Here, due to very strong broad Balmer emission 
lines, not one but three Gaussian functions are applied to describe each broad Balmer 
component. And due to not apparent narrow H$\alpha$ and [N~{\sc ii}] doublet in the spectrum 
in the bright state, the Gaussian components for the narrow H$\alpha$ and [N~{\sc ii}] doublet 
in the bright state have been accepted to have the same central wavelengths and line widths 
as those determined in the dim state. Meanwhile, an additional Gaussian function is applied 
to each emission component of [O~{\sc iii}]$\lambda4959,5007$\AA~ doublet, in order to well 
describe more complicated line profiles of [O~{\sc iii}] doublet in the spectrum whit high 
signal-to-noise in the bright state. Sum of the broadened, shifted and strengthened templates 
of optical Fe~{\sc ii} emission features in \citet{kp10} is applied to describe the optical 
Fe~{\sc ii} emissions in the bright state. Then, through the Levenberg-Marquardt least-squares 
minimization technique, the best fitting results ($\chi^2/dof\sim1.34$) and corresponding 
residuals to the emission lines in the bright state are shown in the middle and right panels 
of Fig.~\ref{bright}. The best fitting results can be applied to confirm that the \obj~ is a 
Type-1 AGN in the bright state, due to strong and apparent broad H$\beta$ and broad H$\alpha$ 
and the flux ratio $4.06_{-0.84}^{+1.01}$ of broad H$\alpha$ to broad H$\beta$. Here, the 
measured narrow emission intensities around H$\alpha$ are only basically consistent with 
those measured in the dim state, mainly due to very weak narrow emission lines relative to 
the very strong broad H$\alpha$ in the bright state.

	Based on the best fitting results above, the continuum luminosities $L_{5100}$ in 
units of ${\rm 10^{43}erg/s}$ at 5100\AA~ in rest frame are $L_{5100,d}=1.16\pm0.06$ and 
$L_{5100,b}=7.78\pm0.06$ in the dim state and in the bright state, respectively. Meanwhile, 
the broad H$\alpha$ luminosities $L_\alpha$ in units of ${\rm 10^{42}erg/s}$ are 
$L_{\alpha,d}=0.19\pm0.02$ and $L_{\alpha,b}=7.94\pm1.02$ in the dim state and in the 
bright state, respectively. Accepted the reported strong linear correlation between 
$L_{5100}$ and $L_{\alpha}$ for SDSS quasars in \citet{gh05}, properties of $L_{5100}$ and 
$L_{\alpha}$ in dim and bright states are checked in \obj~ and shown in top panel of 
Fig.~\ref{cl}. Here, 1158 unobscured SDSS quasars are selected from the database of 
\citet{sh11} with reliable measurements of $L_{5100}$ and $L_{\alpha}$ and with flux ratio 
of broad H$\alpha$ to broad H$\beta$ smaller than 4. It can be confirmed that the \obj~ 
in the dim and bright states follows the same dependence of $L_{\alpha}$ on $L_{5100}$ as 
those for the unobscured SDSS quasars, apparently indicating few effects of dust reddening 
on the measurements of $L_{5100}$ and $L_{\alpha}$ in dim and bright states. Furthermore, 
if accepted the dim state was due to serious obscuration in \obj, due to the estimated 
E(B-V)=1.72 by the extinction curve in \citet{fi99} applied to explain 
$L_{\alpha,b}/L_{\alpha,d}\sim41.8$, the reddening corrected $L_{5100}$ and $L_{\alpha}$ 
from the dim state can lead \obj~ to be an outlier (solid red circle) in the space of 
$L_{5100}$ versus $L_{\alpha}$, to re-confirm few effects of dust obscurations in the dim 
state in \obj.

	Once effects of serious obscurations can be ruled out in \obj, the variability of 
broad emission lines, especially the broad H$\alpha$, should obey the expected results by 
the virialization assumptions in central BLRs. Considering the measured line width $V$ 
(second moment, in units of 1000km/s) of the broad H$\alpha$ of $V_d=0.87\pm0.05$ and 
$V_{b}=2.44\pm0.14$ in dim state and in bright state, the corresponding virial BH masses 
(in units of ${\rm 10^6M_\odot}$) in the dim state and in the bright state are 
$M_d=4.70\pm0.83$ and $M_b=307\pm58$ by the equation 
$\frac{M_{BH}}{\rm M_\odot}=15.6\times10^6(L_\alpha)^{0.55}(V)^{2.06}$ through the equations 
in \citet{gh05} combined with full width of half maximum to be $2.35V$. Here, the 
uncertainties of the virial BH masses are determined by uncertainties of the line parameters 
of corresponding broad H$\alpha$. Meanwhile, through the known M-sigma relation reported 
in \citet{kh13, bt21}, the estimated central BH mass (in units of ${\rm 10^6M_\odot}$) is 
$M_\sigma=217\pm85$, which is simply consistent with the $M_b$ after considering uncertainties. 
Meanwhile, through the selected quiescent galaxies and reverberation mapped broad line AGN 
and tidal disruption events, the M-sigma relation is re-determined as 
$\log(M_{BH}/{\rm M_\odot})=(-2.89\pm0.49)+(4.83\pm0.22)\log(\sigma/{\rm (km/s)})$ by the 
Least Trimmed Squares regression technique \citep{cap13} and shown as solid red line in 
bottom panel of Fig.~\ref{cl}, leading to similar BH mass in \obj. The consistent BH masses 
between $M_\sigma$ and $M_b$ strongly indicate virialization assumptions preferred in the 
BLRs in the bright state of \obj. However, the very different $M_{d}$ from $M_{b}$ apparently 
indicate that the measured parameters of the broad H$\alpha$ cannot follow the expected 
results by the virialization assumptions, i.e., the expected results by the virialization 
assumptions cannot be found in the broad H$\alpha$ in the dim state.


	In order to explain the contrary properties of broad H$\alpha$ to be against the virial 
expected results in the dim state but to do agree with the virial expected results in the 
bright state, rather than the receding dust torus model, but the accretion regulated dust 
torus as discussed in \citet{zh18} can be naturally proposed in \obj. In the dim state, 
weaker accretion rate (smaller continuum luminosity) leading to larger opening angle (relative 
to the equatorial plane, as shown in the toy model in Fig.~\ref{a1} in the Appendix) of central 
dust torus indicates that only far side of central BLRs can be directly observed, but in the 
bright state, higher accretion rate (stronger continuum luminosity) leading to smaller opening 
angle of central dust torus indicates that all the central BLRs can be observed. Therefore, 
in the dim state, the broad H$\alpha$ line luminosity determined BLRs size is totally smaller 
than intrinsic value, leading to viral BH mass very smaller than the M-sigma relation determined 
BH mass. However, in the bright state, the BLRs being common leads the virial BH mass to be 
consistent with the M-sigma relation determined BH mass.

	Meanwhile, as discussed in \citet{zh18}, variations of accretion rates can lead to 
large enough variations in opening angles of central dust torus. In \obj, from dim state 
to bright state, after accepted bolometric luminosity to be 15 times of $L_{5100}$ as recently 
discussed in \citet{nh19}, the dimensionless Eddington ratio is changed from 0.006 to 0.04 
(smaller than the critical value 0.5). Then, the half opening angle $\sim$27\degr of central 
dust torus in bright state could be around 15 degrees smaller than the half opening angle 
$\sim$42\degr in dim state, after accepted the roughly linear dependence shown in Fig.~11~ 
in \citet{zh18}. Certainly, due to large enough scatters in the dependence in 
\citet{zh18}, the estimated half opening angles in \obj~ are not accurate values, but can be 
applied to show the probable variations in half opening angles. Here, it is hard to give an 
quantified structures of central systems in \obj, however, as a toy model in Fig.~\ref{a1} 
in Appendix B, the scenario of accretion regulated dust torus can be reasonably accepted to 
explain the variations of broad emission lines in \obj. 

	Furthermore, we give further discussions to confirm the time duration 13.8 years 
(from the bright state to the dim state) being long enough to complete the regulated process 
of the central dust torus in \obj, by the following two points. First, as discussed in 
\citet{ka22, khr13} on dust torus in the individual AGN NGC 4151, the radius of the dust 
sublimation region could vary over years, through both the near-IR reverberation mapping 
technique and the interferometry results. Second, as the toy model in Fig.~\ref{a1} in 
Appendix, the height variation (length of $\overline{CD}$) about $1.3R_{B}$ can lead moving 
velocity of dust clouds to be $\frac{1.3R_{B}}{\rm 13.8 years}\sim2400{\rm km/s}$, which 
is simply consistent with estimated free fall valocity for dust clouds with 
distance about 150 light-days (4-5times of $R_{B}$) to central BH (BH mass about 
$2-3\times10^8{\rm M_\odot}$) in \obj. Here, the $R_{B}$ can be estimated to be 31.6 
light-days in \obj through the empirical relation in \citet{ben13}, based on the continuum 
luminosity $7.78\times10^{43}{\rm erg/s}$ at 5100\AA~ in the rest frame in the bright state 
in \obj. Therefore, the physical picture on accretion regulated dust torus is efficient 
enough in the CLAGN \obj. 
The results 
above provide clues enough to support that dynamical evolving properties of central dust 
torus have apparent effects on variability properties of broad Balmer emission lines in CLAGN.

\section{Conclusions}

	Motivated by the scenario of radiation regulated dust torus in AGN, variability 
properties of broad emission lines have been checked for testing the evolving dust torus 
in the known CLAGN \obj~ with apparent variations in intrinsic accretion rates. Through the 
strong correlation between $L_{5100}$ and $L_{\alpha}$ for unobscured SDSS quasars, effects 
of moving dust clouds can be ruled out in \obj. Meanwhile, due to the virial BH mass in the 
bright state being consistent with the BH mass estimated by the M-sigma relation, the 
virialization assumptions are efficient enough in the central BLRs in \obj. However, in the 
dim state, the M-sigma relation determined BH mass is very larger than the virial BH mass. 
Therefore, rather than the receding torus model, the scenario of accretion regulated dust 
torus is preferred in \obj, with opening angle of dust torus declining with increasing 
accretion rate below a critical Eddington ratio. In other words, lower accretion rate leading 
to larger half opening angle of central dust torus indicates only outer part of central BLRs 
being detected in the dim state leading to smaller lines widths in broad H$\alpha$, but all 
the BLRs can be detected in the bright state. Therefore, apparent effects of evolving dust 
torus should be expected in studying properties of CLAGN. Furthermore, studying properties 
of broad Balmer emission lines in a sample of CLAGN should provide further clues to determine 
properties of changes in structures of central dust torus in the near future.

\begin{acknowledgements}
Zhang gratefully acknowledge the anonymous referee for giving us constructive comments and 
suggestions to greatly improve the paper. Zhang gratefully thanks the kind grant 
support from the HangJi Action Plan under the Guangxi Science and Technology Program 
2026GXNSFDA00640018 and from NSFC-12373014 and 12173020 and from Guangxi Talent Programme 
(Highland of Innovation Talents) and from the Bagui Scholars Programme (W X G., GXR-6BG2424001). 
This manuscript has made use of the data from the SDSS (\url{https://www.sdss.org/}).
\end{acknowledgements}

\appendix

\section{Emission line parameters}

	The measured emission line properties are listed in Table~\ref{el1}. Based on the 
listed parameters, we can find that there are similar extended components having line width 
(second moment) about 8\AA~ in the [O~{\sc iii}]$\lambda5007$\AA~ in the bright state and 
in the dim state. However, there is an additional narrower component having line width (second 
moment) about 1.7\AA~ in the [O~{\sc iii}]$\lambda5007$\AA~ in the bright state. The narrower 
component was lost in the dim state, probably mainly due to its lower signal-to-noise (13) 
of the spectrum in the dim state than the signal-to-noise (19) of the spectrum in the bright state.

	Furthermore, we can find that the measured narrow H$\beta$ flux is zero in the bright 
state. Therefore, it is necessary to estimate a upper limit $f_{up}$ of narrow H$\beta$ flux 
in the bright state. In other words, narrow H$\beta$ having line flux smaller than $f_{up}$ 
should be not detected in the spectrum in the bright state in \obj. Here, a simple method is 
applied as follows to estimate $f_{up}$. If a narrow H$\beta$ component with line width (second 
moment) 4.6\AA~ (same as the value determined in the dim state) and line flux $f_{line}$ was 
intrinsically included in the spectrum in the bright state of \obj, once such narrow component 
can be detected with its measured parameters at least 1 times larger than their determined 
uncertainties by the same fitting procedure above, we can accept $f_{line}$ as the $f_{up}$, 
leading to $f_{line}=f_{up}\sim30\times10^{-17}{\rm erg/s/cm^2}$. Considering the measured 
narrow H$\beta$ line flux in the dim state is smaller than $f_{up}$, hence such narrow 
component cannot be detected in the spectrum in the bright state.

\begin{table}
\label{el1}
\caption{Limited ranges for the Model parameters}
\begin{tabular}{llll}
\hline\hline
Line & $\lambda_0$ & $\sigma$ & flux  \\
\hline\hline
	\multicolumn{4}{c}{dim state} \\
H$\alpha_B$ & 6565.0$\pm$1.1 & 19.1$\pm$1.2 & 100$\pm$8 \\
H$\beta_N$ & 4860.7$\pm$1.2 & 4.6$\pm$1.2 & 10$\pm$2 \\
H$\alpha_N$ & 6564.1$\pm$0.2 & 2.5$\pm$0.3 & 22$\pm$3 \\
\n2 & 6583.9$\pm$0.2 & 3.5$\pm$0.2 & 39$\pm$3 \\
\o3 & 5000.5$\pm$0.4 & 7.9$\pm$0.4 & 81$\pm$4 \\
\si & 6715.8$\pm$0.7 & 5.1$\pm$0.6 & 19$\pm$2 \\
\sii & 6731.5$\pm$1.4 & 5.1$\pm$0.6 & 14$\pm$2 \\
\hline
\multicolumn{4}{c}{bright state} \\
H$\alpha_{B1}$ & 6562.1$\pm$1.2 & 22.9$\pm$0.7 & 2194$\pm$254 \\
H$\alpha_{B2}$ & 6572.3$\pm$2.4 & 72.9$\pm$4.3 & 1061$\pm$51 \\
H$\alpha_{B3}$ & 6572.8$\pm$1.1 & 15.3$\pm$0.9 & 1081$\pm$251 \\
H$\beta_{B1}$ & 4862.3$\pm$0.8 & 29.1$\pm$1.7 & 580$\pm$50 \\
H$\beta_{B2}$ & 4864.7$\pm$0.3 & 13.1$\pm$0.6 & 488$\pm$54 \\
H$\alpha_N$$^*$ & 6564.1 & 2.5 & 25$\pm$8 \\
H$\beta_N$ & 4860.7 & 4.6 & 0 \\
\o3$_c$ & 5007.2$\pm$0.2 & 1.7$\pm$0.3 & 22$\pm$4 \\
\o3$_e$ & 4999.2$\pm$0.9 & 8.8$\pm$0.7 & 73$\pm$7 \\
\n2$^*$ & 6583.9 & 3.5 & 88$\pm$9 \\
\si & 6718.2$\pm$0.7 & 2.7$\pm$0.8 & 11$\pm$3 \\
\sii & 6732.6$\pm$0.8 & 2.8$\pm$0.8 & 9$\pm$3 \\
\hline\hline
\end{tabular}\\
Notice: The first column shows which emission component is described. The second, the third 
and the fourth columns show the determined central wavelength in units of \AA, the line width 
(second moment) in units of \AA~ and the emission flux in units of $10^{-17}{\rm erg/s/cm^2}$ 
of the corresponding emission component. For the component with suffix $N$ ($B$) means narrow 
(broad) component in Balmer emission line. The narrow emission lines of H$\alpha_N$ and \n2 in 
the bright state are measured by their central wavelengths and line widths to be fixed to the 
values measured in the dim state, therefore, corresponding uncertainties (same as the ones in 
the dim state) are not given.
\end{table}

\section{A toy model on spatial structures of central systems}

\begin{figure}
\centering\includegraphics[width = 9cm,height=5cm]{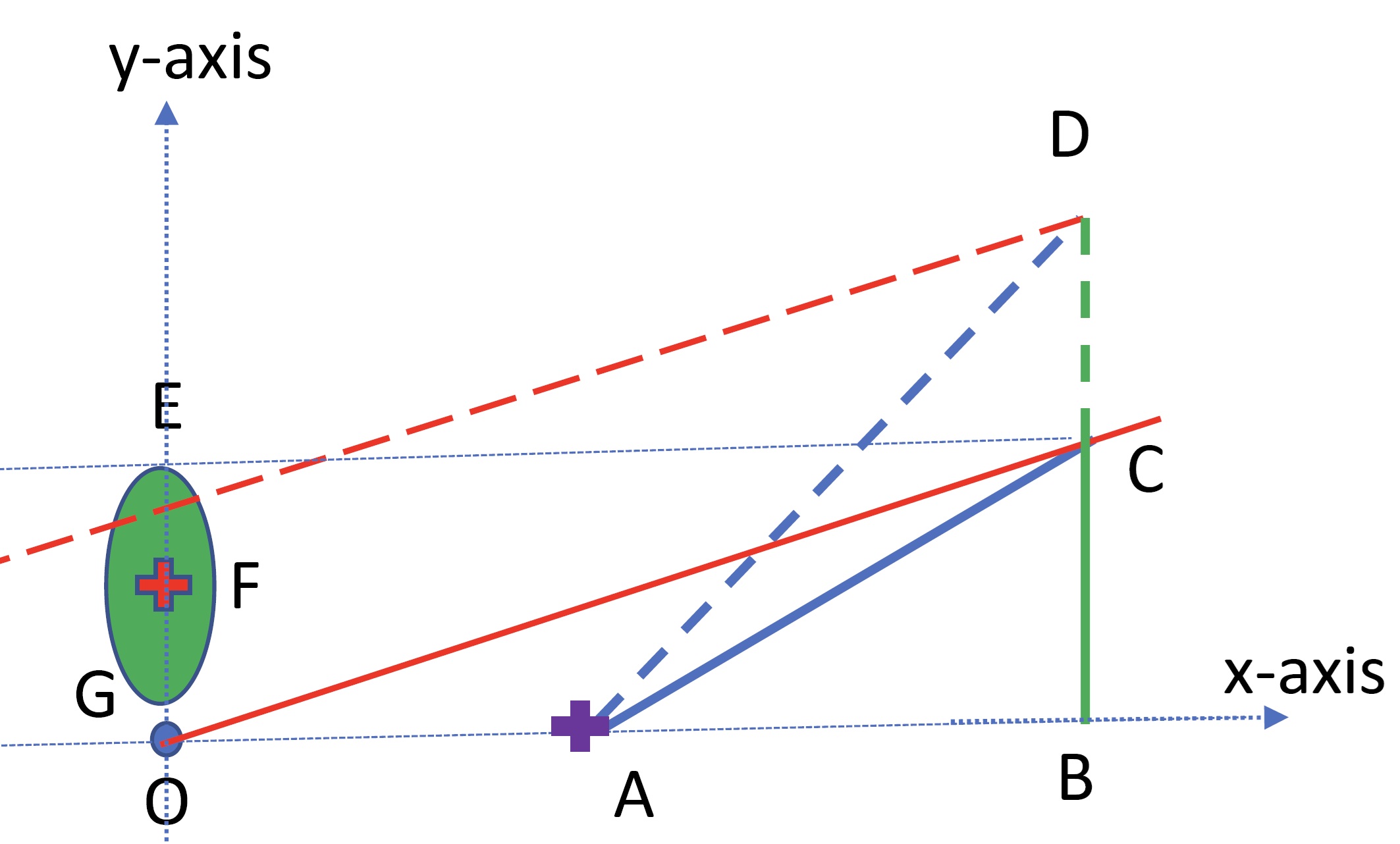}
\caption{Toy model of central systems in \obj. Area in green shows the BLRs with G, F and E 
marking the inner boundary, the central point (red cross) and the outer boundary. Solid 
circle in blue (O) show the central BH. Purple cross (A) shows the inner boundary of central 
dust torus. Lengths of $\overline{BC}$ and $\overline{BD}$ mark the heights of the dust torus 
in the bright state and in the dim state. $\angle BAC$ and $\angle BAD$ represent the half 
opening angles of the dust torus in the bright state and in the dim state. Solid and dashed 
lines in red show the assumed direction of line of sight.}
\label{a1}
\end{figure}

	In order to give more clear descriptions on accretion regulated dust torus in \obj, 
a toy model shown in Fig.~\ref{a1} can be given on spatial structures of central systems. 
The extended spatial structure of BLRs is shown as the area filled by green, with red cross 
(F) as the central point, and with $G$ and $E$ as the inner and outer boundaries of the central 
BLRs. The central BH is shown as solid blue circle (O), lying at the cross point where x-axis 
and y-axis intersect. And the inner boundary (A) of the dust torus is shown as purple cross, 
the upper boundaries of the central dust torus in the bright state and in the dim state are 
shown as solid and dashed blue lines. Then, based on discussions in \citet{km14, sh23}, size 
$R_{D}$ (length of $\overline{OA}$) between central BH and inner boundary of dust torus can 
be simply estimated by bolometric/continuum luminosity, and ratio about 4 of $R_D$ to $R_{B}$ 
(size of BLRs in \obj, length of $\overline{OF}$) can be commonly expected in one AGN. 

	Moreover, based on the $R_{B}$ (size of BLRs) in different states with different 
continuum luminosities in the known reverberation mapped AGN NGC5548 as shown in \citet{pe04}, 
the mean value of $R_{B}$ is about 16light-days, the lower and upper limits of $R_{B}$ are 
about [7, 27]light-days. Therefore, accepted extended structures of BLRs in \obj~ having 
similar ratios as those in NGC5548, the corresponding lower and upper limits of central BLRs 
in \obj~ can be reasonably accepted as $0.4R_{B}$ (length of $\overline{OG}$) and $1.7R_{B}$ 
(length of $\overline{OE}$). Then, we simply accepted that the height of the dust torus shown 
as vertical dashed line in the bright state is equal to the distance ($1.7R_{B}$) between 
outer radius of BLRs to central black hole, in order to ensure that the total BLRs can be 
obscures when line of sight parallel to x-axis.

	Based on the toy model with assumed red lines as line of sight, in bright state, 
areas above the solid red line can be directly observed. If $\angle BAC$ (half opening 
angle in bright state) accepted to be $\sim$27\degr (related to Eddington ratio 0.04) in the 
bright state changed to $\angle BAD\sim$42\degr (related to Eddington ratio 0.006) in \obj, 
the height of $\overline{CD}$ can be estimated to be 
\begin{equation}
	\overline{CD}~=~\overline{AB}\times(\tan(42\degr) - \tan(27\degr))~=~1.3R_{B}
\end{equation}, 
accepted tiny variations of inner boundaries of the central dust torus from the bright state 
to the dim state in \obj. Therefore, from the bright state to the dim state, the region of 
BLRs over the dashed red line (shifted from solid red line with shifted distance $1.3R_{B}$) 
can be directly observed without obscurations, but the region of BLRs under the dashed red 
line could be serious obscured by dust torus. In other words, about 70\% 
($\sim\frac{1.3-0.4}{1.7-0.4}$) of the BLRs can be seriously obscured. Meanwhile, based on 
the shown toy model, the inclination angles of line of sight relative to the x-axis can be 
estimated as $\angle BOC\sim$14\degr, a possible value for Type-1 AGN with high ccretion rates.

	The results above indicate that the assumption of accretion regulated dust torus can 
be reasonably accepted to explain the variations of broad emission lines in \obj. Certainly, 
different values of $\overline{OE}$ ($\overline{BC}$), $\angle BAC$ and $\angle BAD$ can lead 
to different results on obscured regions of BLRs. For example, a large height 
$\overline{OE}$ can sensitively lead to a large value of $\angle BOC$. Unfortunately, at 
current stage, it is hard to give a clear estimations on the parameters in the toy model 
above. Therefore, there are not further discussions on the toy model any more.

	Before ending the section, one point should be noted. As the discussed Failed
Radiatively Accelerated Dusty Outflow (FRADO) models in \citet{na21, na25}, central dust torus 
can be builded related to the FRADO models in AGN. Unfortunately, in cases with low accretion 
rates (SDSS J1011+5442 having lower accretion rates even in bright state), the moving 
velocities of clouds are only a few hundreds of kilometers per second in FRADO related to 
dust torus, which is very smaller than the expected velocities (about 2000${\rm km/s}$) to 
change the height of the dust torus in SDSS J1011+5442. In other words, if accepted the central 
dust torus related to FRADO models, the time duration 13.8 years should be not long enough 
to complete the regulated process of central dust torus in SDSS J1011+5442. Therefore, the 
central dust torus in SDSS J1011+5442 could be probably different from the torus related to 
the FRADO models.

\end{document}